# Multifunctional Biocomposites based on Polyhydroxyalkanoate and Graphene/Carbon-Nanofiber Hybrids for Electrical and Thermal Applications

*Pietro Cataldi\*, Pietro Steiner, Thomas Raine, Kailing Lin, Coskun Kocabas, Robert J. Young, Mark Bissett, Ian A. Kinloch\*, Dimitrios G. Papageorgiou\*[†]*

Department of Materials and National Graphene Institute, University of Manchester, Oxford Road, Manchester, M13 9PL UK
[†] School of Engineering and Materials Science, Queen Mary University of London, Mile End Road, London E1 4NS, UK

\* Corresponding authors e-mails: pietro.cataldi@manchester.ac.uk, d.papageorgiou@qmul.ac.uk, ian.kinloch@manchester.ac.uk

Abstract: Most polymers are long-lasting and produced from monomers derived from fossil fuel sources. Bio-based and/or biodegradable plastics have been proposed as a sustainable alternative. Amongst those available, polyhydroxyalkanoate (PHA) shows great potential across a large variety of applications but is currently limited to packaging, cosmetics and tissue engineering due to its relatively poor physical properties. An expansion of its uses can be accomplished by developing nanocomposites where PHAs are used as the polymer matrix. Herein, a PHA biopolyester was melt blended with graphene nanoplatelets (GNPs) or with a 1:1 hybrid mixture of GNPs and carbon nanofibers (CNFs). The resulting nanocomposites exhibited enhanced thermal stability while their Young's modulus roughly doubled compared to pure PHA. The hybrid nanocomposites percolated electrically at lower nanofiller loadings (7.5 wt.%), compared to the GNP-PHA system (10 wt.%). The electrical conductivity at 15 wt.% loading was ~ 6 times higher than the GNP-based sample. As a result, the electromagnetic interference shielding performance of the hybrid material was around 50% better than the pure GNPs nanocomposites, exhibiting shielding effectiveness above 20 dB, which is the threshold for common commercial applications. The thermal conductivity increased significantly for both types of bio-nanocomposites and reached values around 5 W K$^{-1}$ m$^{-1}$ with the hybrid-based material displaying the best performance. Considering the solvent-free and industrially compatible production method, the proposed multifunctional materials are promising to expand the range of application of PHAs and increase the environmental sustainability of the plastic and plastic electronics industry.

## 1. Introduction

Plastics are fundamental materials in sectors as diverse as packaging, construction, automotive, electronics, medicine, and sports. Today, a world without synthetic polymers appears unthinkable.[1] Their massive success is driven by their low cost, diverse properties, low maintenance, prolonged lifetime and versatile manufacturing.[2, 3] Most of the existing polymers are constituted by monomers derived from oil, coal or gas.[4] Due to the non-renewable nature of these resources and the longevity of the synthetic plastics, the urge for green alternatives is becoming a priority.[5] In addition, plastics waste management is highly complex considering that several million tons of polymer debris are burned after disposal, producing toxic gases[6] and, as such, contribute to environmental pollution when disposed incorrectly.[7, 8] As a result, bio-based and/or biodegradable plastics, were proposed as a sustainable alternative to synthetic and long-lasting polymers.[3, 9, 10]

Bio-based polymers extracted from renewable biomass sources (e.g. cellulose, starches, sugars from cane and beats, etc.) can reduce the use of fossil fuels.[10] Additionally, biodegradable plastics can degrade in the environment and thus reduce environmental pollution, also contributing positively to the ubiquitous microplastic problem.[10-12] Among them, the most promising to replace fossil-derived polymers are thermoplastics such as thermoplastic starches (TPS) and polylactic acid (PLA) because of their melt processability. The TPS are bio-based and biodegradable but have a low melting temperature (≈ 60 ºC), high vapor permeability, and poor mechanical properties that are often improved by blending with other polymers.[13, 14] On the other hand, PLA displays a higher melting temperature (≈ 150 - 170 ºC)[15, 16] and better mechanical characteristics but is not biodegradable and needs specialized facilities for composting.[17, 18] There is therefore still a strong need to expand the use of thermoplastic bio-based and/or biodegradable biopolymers.

Polyhydroxyalkanoate (PHA) is a thermoplastic biopolymer produced by the fermentation of glucose-rich materials, volatile fatty acids or organic waste, performed by bacteria and exemplifies a great opportunity to boost the applications of biopolyesters.[19-21] It is bio-based and biodegradable and given the fact that more than 150 different PHA monomers have been documented, it displays a huge potential to diversify into a large variety of physically and chemically different types of biopolymers, depending on the combination of monomer units.[22-24] For example, by changing the monomer composition, the melting temperature of PHAs can vary between 40 to 180 °C, the glass transition temperatures can range from -50 to 4 °C and significant variations have been measured also in terms of the degradation temperature, oxygen transmission rate, Young's modulus and tensile strength.[25-27] Furthermore, PHAs are resistant to ultraviolet light and can be employed in piezoelectric applications.[25, 27] The most popular type of PHA is polyhydroxybutyrate (PHB) which is the simplest (i.e. a single monomer). It is, however, a mechanically weak form of this biopolyester and has manufacturing limitations since its thermal degradation begins a few degrees above its melting temperature.[23, 28-30] Although some companies recently launched innovative PHA-based polymer grades that can also reduce the overall cost of the biopolymer,[31, 32] due to its biocompatibility and non-toxicity and its relatively poor thermal and mechanical properties, PHA-using applications are mostly limited to the cosmetic, packaging, drug delivery, and tissue engineering sectors.[25, 33-37]

An expansion of the applications of PHAs can be achieved by developing composites/nanocomposites where PHAs are used as the polymer matrix.[38] As a result of this procedure, some of the drawbacks of PHAs can be tackled and multifunctional features can be acquired. The most popular strategies to realize composites/nanocomposites based on PHA relies on coupling it with other polymers/biopolymers,[39-42] fibers,[43-45] nanoclay,[41, 46-48] and nanocellulose[49-51] with the prevalent intention of improving their mechanical behaviour, gas barrier properties, and thermal stability. Other promising nanofillers are carbon-based ones that show the ability to enhance the intrinsic properties of polymers and, at the same time, add new functionalities to a matrix especially when hybrid nanofillers are included, to take advantage of any additive or synergistic effects.[52] Amongst them, carbon nanotubes (CNTs) have been the most utilized and were mostly solvent mixed with PHAs targeting applications in tissue engineering.[53-60] A few studies reported melt blending of PHA with CNTs[61, 62] and some of the targeted applications of these nanocomposites included conductors for electronics.[61, 63, 64]

Other carbon-based nanomaterials such as graphene-related materials (GRMs) or carbon nanofibers (CNFs) have been less commonly used to boost the properties of PHAs although these nanofillers are nowadays produced in large scale and are usually cheaper than CNTs. Amongst GRMs, graphene nanoplatelets (GNPs) are an ideal candidate for the polymer industry in light of their low price (less than 0.1 $/g compared to a few $/g for CNFs), the large volume of production and competitive performances.[65] When GRMs or CNFs were coupled with PHAs, solvent processing was the most common blending technique [66-75] for the production of nanocomposites targeting biomedical applications [68]. As a result, there is still the need for further enhancing the properties of PHAs and expanding the applications of these biopolymers into fields such as electronics[67, 75] or thermal dissipation. Another important breakthrough will be to demonstrate that a simple, scalable, solvent-free and eco-friendly manufacturing processes can be used to manufacture PHAs nanocomposites and that nanofillers such as GNPs and CNFs are a good candidate to fill the performance gap between PHAs and other biopolymers/polymers.

Herein, melt processing was employed to mix PHAs with GNPs or GNPs-CNFs hybrids (ratio 1:1) to obtain multifunctional materials. GNPs were preferred compared to CNFs since they are high-performance, cheap and multifunctional nanofillers easily produced on a large scale. Both pure GNPs and hybrids nanofillers were homogeneously dispersed inside the biopolymer matrix and increased the thermal stability of the nanocomposite. The Young's modulus of the nanocomposites was measured trough stress-strain tests and was greatly improved with the inclusion of both GNPs and the hybrid nanofillers. The PHA matrix turned from an electrical insulator to a conductor with nanofiller inclusion, with the GNPs-CNFs biocomposite showing impedance measurements giving electrical conductivity higher than the GNP-based sample. We tested the electromagnetic interference (EMI) performance of the composites to demonstrate application related to their electrical conduction. The thermal conductivity increased significantly for both types of bio-nanocomposites, with a maximum increase of almost 2000% compared to the pure PHA matrix. The biocomposites obtained could expand the applications of PHA-based materials into electronics as for example as EMI shielder, thermally dissipative and structural materials and at the same time increase the environmental sustainability of the plastic industry.

## 2. Results and Discussion

2.1 Enhanced Thermal Stability and Morphology

The production of nanocomposites by melt mixing meant that the use of solvents could be avoided, improving the eco-friendly nature of the material. The nanocomposites contained either GNPs or a hybrid mixture of GNPs and CNFs at a ratio of 1:1, with the concentrations of the fillers ranging from 2.5 wt.% to 15 wt.%. GNPs were selected since they are high-performing, multifunctional nanofillers. Their properties might not be comparable with other carbon-based nanofillers but GNPs can be easily produced on a large scale and therefore are much cheaper compared with CNTs and CNFs. The PHA-based nanocomposites were fabricated by internal mixing and compression molding at 130 °C (see materials and methods section for more details). Thermogravimetric analysis (TGA) revealed a degradation temperature of ~ 275°C in nitrogen for the pure PHA, while the inclusion of nanofillers had a positive effect on the thermal stability of the nanocomposites as shown in Figure S1. Pure PHA has a $T_{50\%}$ of 276 ± 2 °C while the addition of 15 wt.% nanofillers in the form of GNPs or hybrid GNPs-CNFs increases $T_{50\%}$ to 289.4 ± 0.9 °C and 287 ± 1 °C, respectively. This indicates that both GNPs and CNFs provide a substantial increase in the thermal stability of the nanocomposites. Generally, the higher loadings of GNPs provide superior thermal stability compared to the hybrids, suggesting that higher loadings of high aspect ratio materials provide a greater barrier to the diffusion of degradation products.[76, 77]

Pristine PHA samples macroscopically were rigid and pale yellow in color (inset of Figure 1a) with smooth and uniform surface topographies at the micrometer scale, as displayed in Figure S2. The thickness of all the samples was about 1 mm (see Figure S3). The neat PHA polymer exhibited a regular cross-section with no air bubbles/holes visible through the section (Figures 1a S3). The nanofillers were homogeneously dispersed within the polymer matrix (Figures 1b and 1c). The GNPs-based nanocomposites exhibited the roughest surface (Figure 1b) with roughness increasing with GNPs loading (Figure S3) and a higher amount of restacked flakes and agglomerates.[78] The presence of CNFs can be identified in the hybrid sample presented in Figure 1c, (see Figure S3), while the distribution of both fillers seems to be more homogeneous. The polymer-filler interface for the majority of the flakes in both samples appeared to be intact, without the presence of any gaps or voids; this fact is expected to maximize the reinforcement efficiency. Additionally, a small number of flakes can be seen to have folded as a result of the melt-mixing procedure within the internal mixer. This phenomenon is commonly observed when GNPs are mixed with low-modulus elastomers under high shear rates.[79, 80] Finally, it is interesting to observe that the compression molding procedure that took place after mixing, contributed to a preferred orientation to both the GNPs and the CNFs in the axial direction of the samples. In this case, even if the degree of orientation is not very high, it is expected to contribute to enhanced in-plane mechanical properties and activation of the conduction mechanisms.

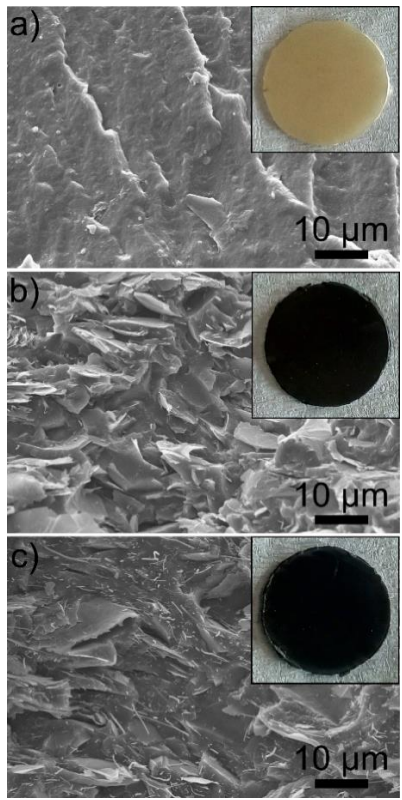

Figure 1: a), b) and c) display the cross-section SEM image of the pure PHA, of the 15 wt.% GNPs sample and of the 15 wt.% hybrid, respectively. Insets shows the photograph of the samples (diameter of the disk around 2.5 cm).

2.2 Mechanical Properties

The mechanical properties of the samples were evaluated by tensile testing and the resulting stress-strain curves are shown in Figure 2a,b. The presence of the nanofillers greatly enhanced the modulus of the matrix by almost doubling it at the highest filler content (15 wt.% or ~ 9 vol%) (Figure 2c,d). The samples were labeled as $x$GNP or $x$Hyb were $x$ indicates the weight percent of the nanofillers inside the PHA matrix and GNP or Hyb indicate if the nanofiller was pure nanoflakes or the hybrids of GNPs and CNFs, respectively. The super-linear increase of the modulus of the nanocomposites at loadings higher than 5 wt.% (3 vol%) indicates that the fillers exert an additional enhancement at higher filler contents, as a result of the simultaneous contribution from both individual nanoplatelets and pair of fillers.[79] The two types of fillers seem to have acted additively towards the improvement of the mechanical properties of the bio-nanocomposites similar to what we have seen in hybrid composites where GNPs were combined with carbon fibres[81] and glass fibres[82]. Regarding the tensile strength, it can be seen from Figure 2d that the combination of both CNFs and GNPs hardly altered the behavior of the nanocomposites, while the presence of only GNPs reduced the strength of the matrix. Given that the tensile strength is generally more sensitive to aggregation compared to the modulus of polymer nanocomposites[83], it can be concluded that the samples filled with only GNPs displayed a higher amount of aggregates compared to the hybrid samples. Finally, as expected, the presence of the nanofillers in both types of composites reduced the elongation to failure of the PHA matrix.

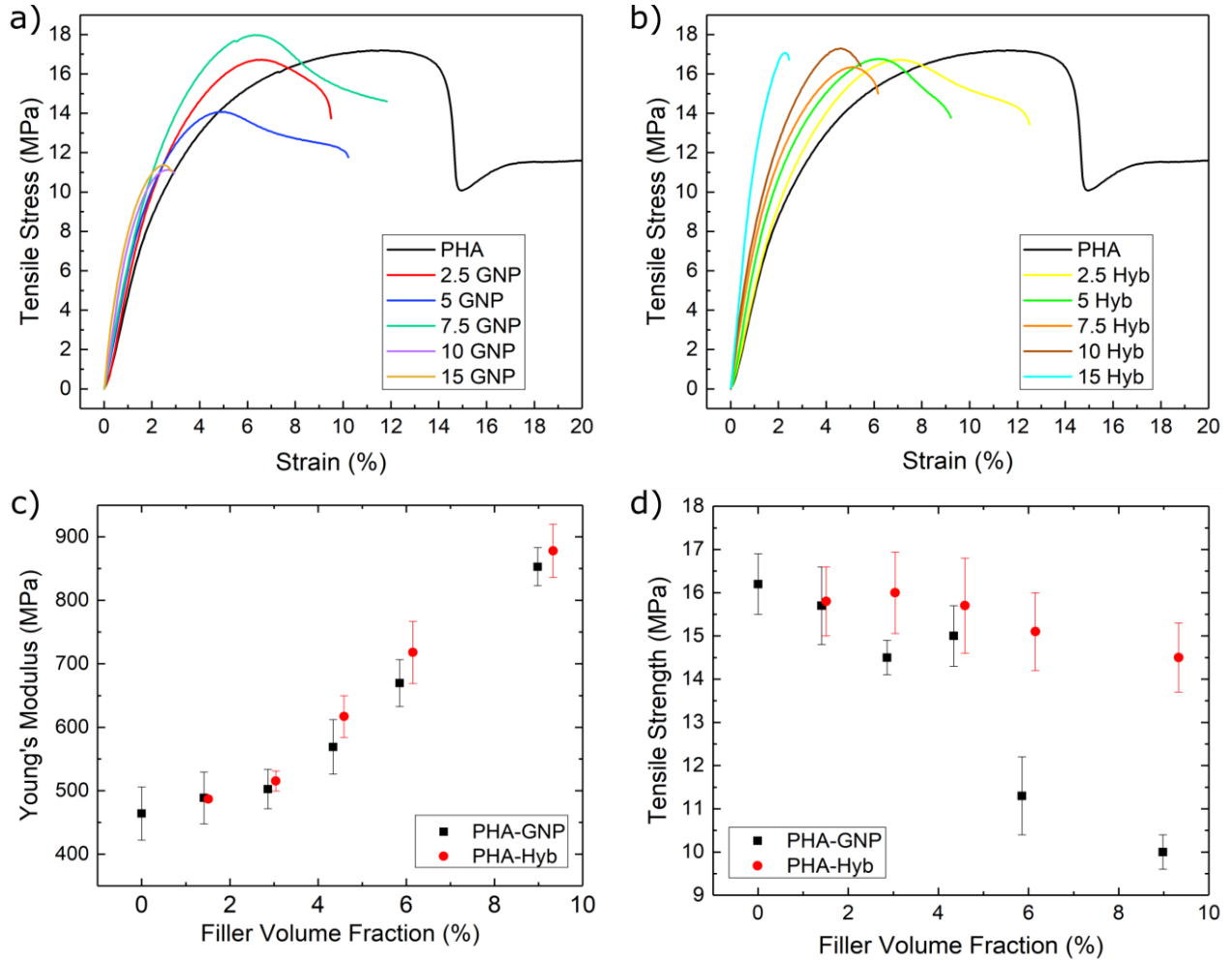

Figure 2: Stress-strain curves of PHA samples filled with a) GNPs and b) Hybrid (GNP-CnF) nanofillers. The results of c) the Young's modulus and d) tensile strength for both types of bio-nanocomposites

2.3 Electrical conductivity

The inclusion of conductive nanofillers can increase the electrical conduction of polymer composites.[84-88] Figures 3a-b show the specific electrical conductivity as a function of the frequency of the input current for the GNP- and Hyb-based samples, respectively. Low conductivities defining an insulating behavior, were measured for the pure PHA biopolymer and for the GNP-based samples up to 7.5 wt.% loadings (Figure 3a). In contrast, the specific electrical conductivity increased by about nine orders of magnitude at 10 wt.% GNPs loading.

This suggests that the electrical percolation threshold, defined as the loading at which the conductive nanofillers form efficiently interconnected conductive pathways,[89] takes place between 7.5 and 10 wt.% loading of GNPs. After percolation, the electrical conductivity increased, reaching values of ~ 0.1 S/m and ~ 0.3 S/m for nanocomposites filled with 10 wt.% and 15 wt.% of GNPs, respectively. A lower percolation threshold was detected for the hybrid-based nanocomposites, as shown in Figure 3b. Indeed, the electrical conductivity improved of five orders of magnitude at 1 Hz frequency when the concentration of the hybrid fillers increased from 5 wt.% to 7.5 wt.%, suggesting that percolation occurred in between these two loadings. After percolation, the electrical conductivity increased with increasing filler content, reaching values of ~ 0.1 S/m and ~ 2 S/m for 10 wt.% and 15 wt.% loadings, respectively. It is worth noticing that at the highest nanofillers loading, the Hyb-based biocomposite displays an electrical conductivity that is ~ 6 times higher than the GNP-based sample at the same loading. This is due to the higher number of connections between nanofiber-nanoflakes in the hybrids samples compared with the pure GNPs specimen.[52, 65] The conductivities obtained reveal the possibility of exploitation of these eco-friendly materials in applications such as electromagnetic interference shielding (seen later in Figure 4), structural health monitoring[90] and electrodes for tactile/pressure sensors [65, 91].

Both pure GNP and Hyb biocomposites showed a two-phase capacitive and/or resistive structure behavior.[92] Before the percolation threshold, a dielectric behavior was measured with the conductivity being linearly dependent on the frequency (capacitive state). In contrast, above percolation the electrical conductivity remained constant with increasing frequency, determining the formation of a conductive network in the biopolymer matrix (resistive behavior). Where the conductivity exhibited a linear dependence on frequency the predominant mechanism was capacitive behavior for the low filler loadings (i.e. before electrical percolation). The prevalence of the resistive component at high filler loading implies the formation of effective percolation networks. The 7.5 wt.% Hyb sample displayed resistive performance at low frequencies (with electrical conductivity independent on frequency) and a switch to capacitive behavior at frequencies around $10^5$ Hz, indicating that the conductive networks are not yet fully formed at this filler loading.

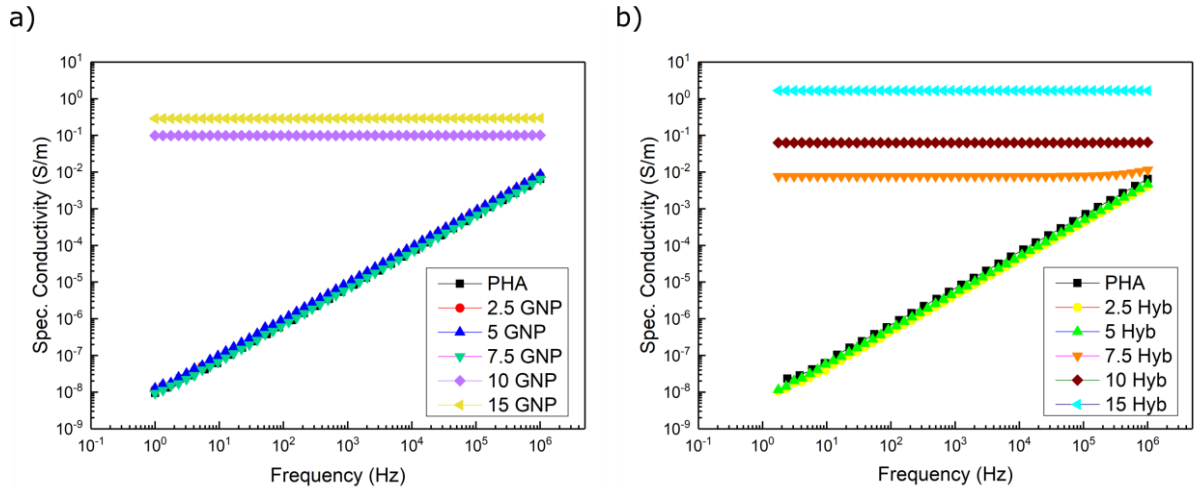

Figure 3: Graph of electrical conductivity as a function of frequency for GNP (a) and (b) Hyb (1:1 GNP to CNF ratio) samples.

2.4 Biopolymer-based Electromagnetic Interference Shielding

Electrical devices are generally designed as arrays of densely packed electronic components.[84] As a result, the generation of EMI can lead to failure of the expected operation of electrical instruments caused by the crosstalk of adjacent electronic units.[93] Consequently, effective EMI shielding is crucial to safeguard the optimal performance of electronics.[94] Metals are commonly used for EMI shielding although they are mechanically rigid, corrosion-prone and expensive.[95] Recently, conductive nanofiller-reinforced biopolymers have been suggested as an eco-friendly alternative to eliminate and control EMI.[84, 95] Herein, we tested the EMI shielding performance of the GNP- and Hyb-based biocomposites as a function of nanofiller content. The EMI shielding was measured at frequencies between 8 and 12 GHz (X-band), which are the most common frequencies for the operation of consumer electronics.[96] During EMI shielding an electromagnetic wave is reflected, absorbed and transmitted when it passes through a conductor. The larger the electromagnetic screening, the lower the amplitude of the transmitted wave. The EMI shielding effectiveness (*SE*) is calculated by:

$$SE \text{ (dB)} = -10 \log(T) \tag{1}$$

where *T* is the transmittance and represents the ratio between the transmitted and the incident electromagnetic power and is normally dependent on the frequency of the incoming electromagnetic wave.[84, 96]

Figure 4a displays the transmittance as a function of the frequency of the incident electromagnetic waves for GNP- based samples with different nanoflake concentrations. As can be seen, increasing the nanofiller concentration enhances the electromagnetic shielding effect. The screening efficiency was found to improve of approximately -1.04 dB per wt.% of GNPs, as shown in Figure S4. The 10 and 15 wt.% GNP-based samples revealed a transmittance of approximately -11.2 and -15.4 dB, respectively. The inclusion of the nanofillers produced an enhancement of the SE also for the composites containing the hybrid fillers, with an improvement of -1.4 dB per unit wt.% of nanofillers, as shown in Figure S4. The 10 and 15 wt.% Hyb-based samples showed a transmittance of approximately -14.6 and -22.7 dB, respectively. These values are approximately 30% and 50% higher compared to the GNP-based samples loaded with the same filler amounts. Considering that the thickness of the samples was constant (1 mm), the Hyb-based biocomposites were able to provide better EMI shielding compared to the ones reinforced with just GNPs, giving 0.023 dB/μm against 0.015 dB/μm at the highest filler loading, respectively. This result is in agreement with the lower electrical resistance that the Hyb samples showed compared to the composites with GNPs, as shown in Figure 3. It is worth noticing that the 15 wt.% Hyb samples display shielding effectiveness higher than 20 dB, which is the threshold for numerous commercial applications.[84, 95, 96]

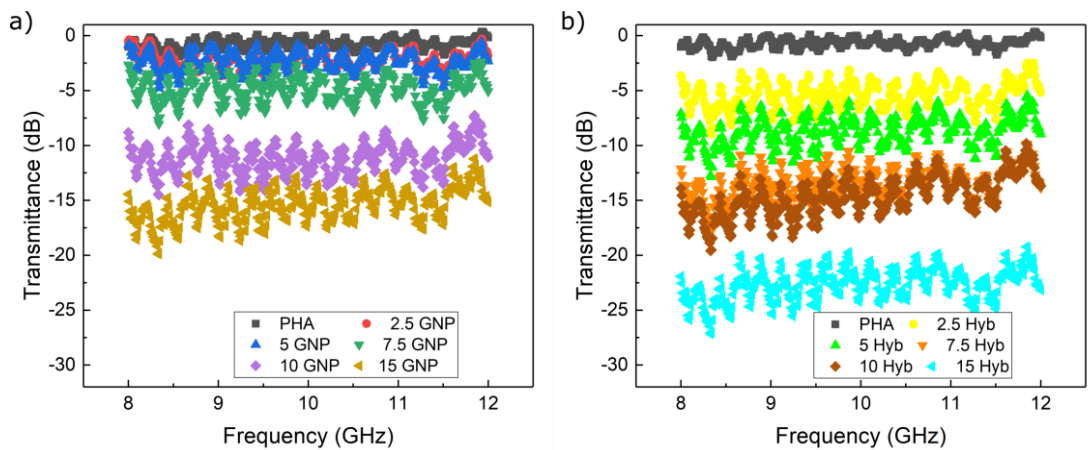

Figure 4: Transmittance of the biocomposites between 8 and 12 GHz. a) and b) GNP- and hybrid-based composites, respectively.

2.5 Enhanced Thermal Dissipation Properties

Another important issue in electronic devices is the heat generated by their resistive components due to the Joule effect. Such heating effects are increasing due to the constantly growing density of microprocessors of personal computers and smartphones (according to the trend predicted by Moore's law) and the resulting enlarged power consumption of the electronics components.[97] Often, thermal dissipation concerns are coupled with EMI shielding dysfunction due to the abovementioned proximity of electrically conductive elements and their interference.[93] The possibility to solve both problems at once by employing eco-friendly materials could constitute a great advantage for electronics manufacturers.[93, 98] Hence, we measured the thermal diffusivity (TD), the rate at which one material is able to transfer heat, and we extracted the thermal conductivity of the GNP- and Hyb- based biopolymers, as shown in Figure 5. The neat PHA matrix displayed a TD of 0.15 mm$^2$/s (Figure 5a), a quite typical value for the majority of polymers due to their amorphous structure.[99] At low filler content, the fillers are located at a distance from one another, not forming a percolating network. With increasing loading of GNPs, the thermal diffusivity continued to increase, almost exponentially, reaching values of 0.53 and 1.16 mm$^2$/s at 10 and 15 wt.% nanoflakes concentrations, approximately 250% and 670% higher than the bare biopolymer, respectively. The inclusion of the hybrid nanofillers resulted in an even more evident improvement of the thermal diffusivity, as shown in Figure 5b. Indeed, at 10 and 15 Hyb samples TD reached values of 0.75 and 1.38 mm$^2$/s, an increase of 500% and 920%, respectively. The diffusivity is greatly enhanced with increasing mass fractions of the nanofillers but especially at 15 wt.% (around 9 vol%) of either GNPs or the Hyb filler, as a result of the creation of thermally conductive pathways inside the biopolymer. Especially for the case of the hybrid 1D-2D nanofillers, the carbon nanofibers with their exceptionally high aspect ratio are able to bridge the GNPs, thus creating highly conductive graphene-graphene networks. Regarding the thermal conductivity of the samples, the increase of thermal conductivity was impressive for both types of composites, with the rate of increase being exponential and the maximum conductivity at the highest filler contents (around 9 vol%) being 1750% higher for the GNP-filled PHA and 1930% higher for the Hyb-filled PHA. These values are in the same range of the vast majority of commercially available thermal interface materials, which display thermal conductivities in the range of 0.5-10 Wm$^{-1}$K$^{-1}$, generally achieved at very high filler volume fractions of around 50 vol%[100] (that is almost 5 times higher than the filler contents used in this work). These values are comparable with state-of-the-art thermal conductivity obtained using other carbon-based nanofillers inside the polymer matrix (see Table 1).

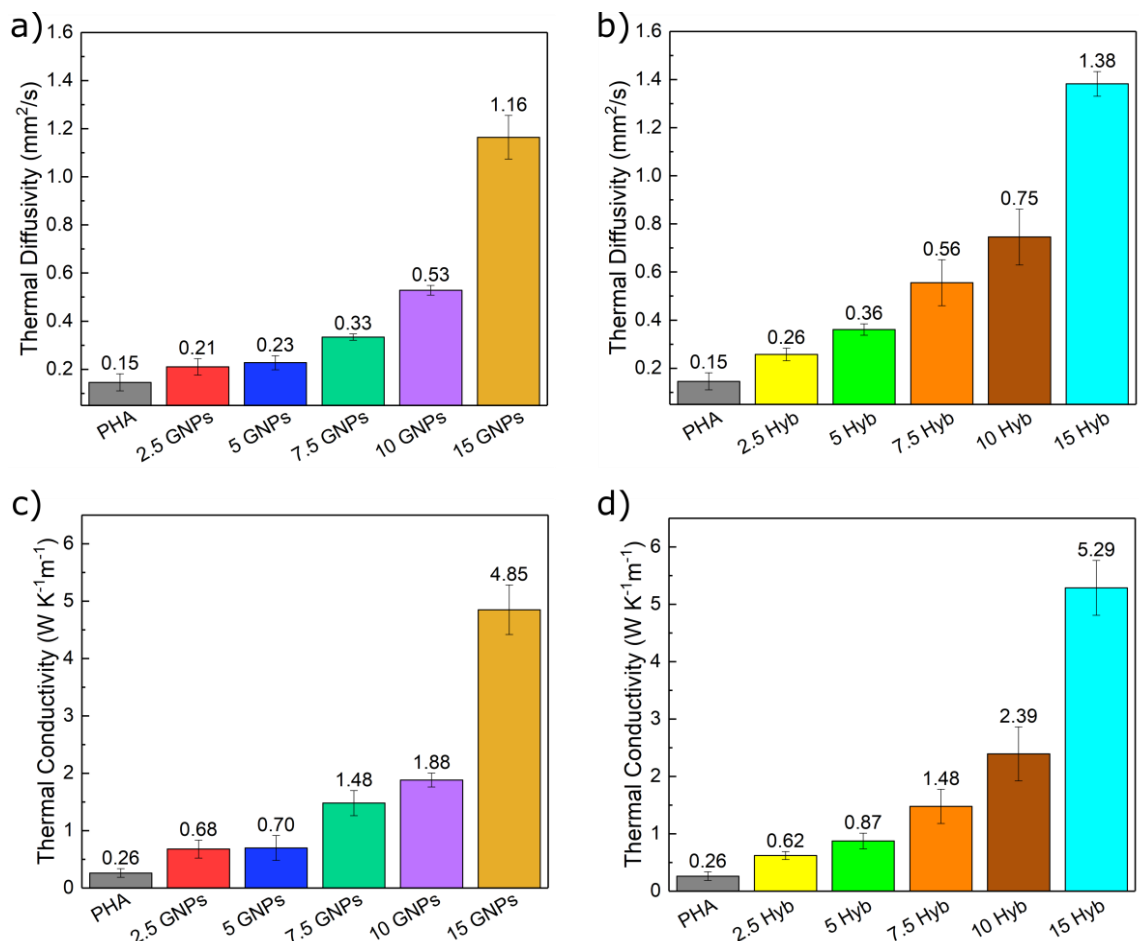

Figure 5: Thermal diffusivity of the bio-nanocomposites filled with (a) GNPs and (b) hybrid, binary GNP-CnF fillers as a function of nanofiller weight fraction (wt.%). Thermal conductivity of the nanocomposites filled with (c) GNPs and (d) hybrid GNP-CnF (Hyb) fillers.

Table 1: Thermal conductivity of selected nanocomposites that employ carbon-based nanofillers.

| Sample | Nanofiller load (vol.%) | Method | Thermal Conductivity (W K$^{-1}$ m$^{-1}$) | Reference |
|---|---|---|---|---|
| GNP-epoxy | 6 | Infrared thermography | 0.6 | [101] |
| GNP-epoxy | 25 | Steady state method | 6.9 | [102] |
| GNP-Epoxy | 10 | Laser flash | 5.1 | [100] |
| GNP/CNT-epoxy | 0.6 | Hot disk | 0.3 | [103] |
| GNP-epoxy | 24 | Steady state method | 6.5 | 104 |
| GNP-epoxy | 45 | Hot disk | 11.0 | 105 |
| GNP-epoxy | 1.2 | Laser flash | 0.5 | 106 |
| CNF-rubbery epoxy | 31.6 | Hot disk | 1.9 | 107 |
| GNP-polycarbonate | 12 | Hot disk | 7.3 | 108 |
| CNT-polydimethylsiloxane | 1.8 | ASTM method | 1.8 | 109 |
| GNP/CNF-PHA | 9.0 | Angstrom Method | 5.3 ± 0.3 | This work |

## 3. Conclusions

In pursuit of enhancing its features and achieving multifunctionality, a PHA biopolyester was melt blended with GNPs or hybrids of GNPs and CNFs. The inclusion of nanofillers increased the thermal stability of the nanocomposites by ~10°C. At the highest nanoparticle loading (15 wt.%), the Young's modulus of the matrix roughly doubled while the tensile stress decreased only around 10% and 40% for the hybrids and GNPs reinforced samples, respectively. The hybrid nanocomposites displayed electrical percolation at filler content between 5 and 7.5 wt.%, whereas the pure GNP-based samples became conductive at loadings between 7.5 and 10 wt.%. The electrical conductivity of the biopolymer increased considerably, with the best performance obtained using 15 wt.% of the hybrid filler that was ~ 6 times higher compared to the GNP-based sample at the same loading. As a result, the EMI shielding performance of the hybrid nanocomposites was 50% higher than the GNPs samples and displayed shielding effectiveness > 20 dB, which is the threshold for commercial applications. The thermal conductivity increased significantly for both types of bio-nanocomposites and reached values around 5 W K$^{-1}$ m$^{-1}$, with the maximum conductivity at the highest filler content being 1750% and 1930% higher than the pure PHA matrix in for the pure GNP- and hybrid-based nanocomposites case, respectively. The multifunctional materials developed show very promising multifunctional properties, they can expand the range of application of PHA biopolymers and increase the environmental sustainability of the plastics industry.

## 4. Materials and Methods

PHA was acquired from Goodfellow (grade PH326302). GNPs were obtained from XG Sciences (grade M25) and were fully characterized (lateral size, Raman spectra) in a previous report[110]. Graphitized CNF (diameter ≈ 100 nm, length between 20 and 200 µm) was purchased from Sigma‐Aldrich (grade PR‐25‐XT‐HHT from Pyrograf Products Inc.) and characterized here earlier.[91]

The melt mixing of the nanocomposites was undertaken in an internal mixer (Thermo Fisher HAAKE Rheomix) at 130 °C and at 50 rpm for 7 min. The GNP fractions of the PHA-GNP set of samples was 2.5, 5, 7.5, 10 and 15 wt.%, while the same fractions were used for the PHA-Hyb set of samples, where the ratio between the GNPs and the carbon nanofibers was 1:1 by weight. Thin films of about 1 mm thickness were created by compression molding the nanocomposites at 130 °C under a pressure of 30 bar for 5 minutes.

The stress-strain curves of the samples were obtained using dumbbell-shaped specimens in an Instron 4301 machine, using a load cell of 5 kN and under a tensile rate of 50 mm min$^{-1}$.

Thermal gravimetric analysis was carried out on TA Instruments Q5500 TGA. The data was obtained by averaging triplicate samples ran under a nitrogen atmosphere at a heating rate of 10°C min$^{-1}$ to 800 °C.

SEM images of the morphology and of the cross section of the specimens were obtained with a Zeiss Evo50 microscope (acceleration voltage of 10 kV). For cross-sectional SEM pictures, the samples were fractured after being frozen in liquid nitrogen.

The impedance of the bio-nanocomposites were measured on specimens of 10 mm × 10 mm × 1 mm in in-plane direction. All the measurement were undertaken on a PSM 1735 Frequency Response Analyzer from Newtons4th Ltd connected with Impedance Analysis Interface (IAI). The frequency of alternating current for the measurement were at a range of 1 to 10$^6$ Hz. The specific conductivity ($\sigma$) of composites were calculated from the impedance by the equation:

$$\sigma(\omega) = \frac{1}{Z^*} \times \frac{t}{A} \qquad (3)$$

Where $Z^*$ is the complex impedance, $t$ is the thickness of specimens (1 mm in this case), $A$ is the area of specimens (10 mm × 10 mm).

The EMI shielding effectiveness of the samples was tested using a vector network analyser (Keysight N5227A) and two WR-90 (8.2-12.4 GHz) waveguide. The transmittance was measured between 8 and 12 GHz.

The in-plane thermal diffusivity of the samples were measured by a custom-built system (scheme of the measuring setup in Figure 6a) using a pulsed (1 Hz) tuneable laser beam and high resolution infrared camera (FLIR T660) mounted with a IR micro lens with a resolution of 50 µm. The laser was used to locally generate periodic heat waves that propagate in the material and are recorded by the infrared camera (Figure 6 b-e). Thermal diffusivity is then calculated following the Angstrom method.[111] For each sample eight measurements were performed along the X and Y axis. A detailed description of the Angstrom method is given in the supporting information (see figure S5).

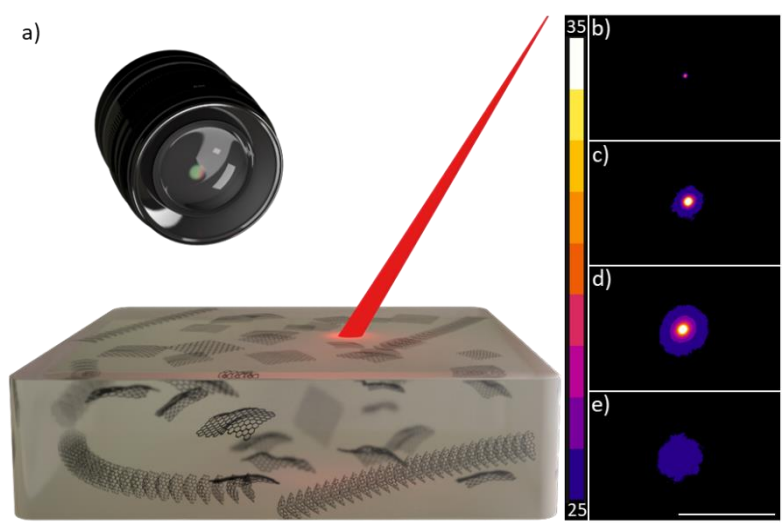

Figure 6: a) schematic representation of the custom-built setup for measuring the thermal diffusivity. The laser is used to produce a periodic heat waves that are recorded by the Infrared camera. b-e) four recorded IR photograms highlighting the propagation of the heat wave into the sample, scale bar 1 cm.

Heat capacities were obtained from modulated DSC runs on a TA Instruments Q100 DSC for samples weighing 3 to 6 mg, ran in triplicate and averaged. The heat capacity at 25 °C was extracted using TA Universal Analysis software. The samples were heated from -50 to 180 °C at 3°C min$^{-1}$ with a modulation amplitude of 0.5°C min$^{-1}$ and a period of 60 s. The heat capacities ($C_p$) and the densities ($\rho$) were used to convert the thermal diffusivity ($T_D$) into thermal conductivity ($\kappa$) using equation:

$$T_D = \frac{K}{\rho C_p} \qquad (2).$$

All the above mentioned measurements were performed on at least three specimens unless specified differently.

Acknowledgements

This project has received funding from the European Union's Horizon 2020 research and innovation programme under grant agreement No 785219.

**TOC**

Bio-based and/or biodegradable plastics have been proposed as a sustainable alternative to long-lasting fossil derived ones. Polyhydroxyalkanoate (PHA) displays great potential across a large range of applications but exhibits poor electrical and physical properties. Here, PHA is melt blended with carbon-based nanofillers, resulting in boosted mechanical, electrical and thermal features. Considering the solvent-free and industrially compatible production method, the proposed multifunctional materials are promising to expand the range of application of PHAs.

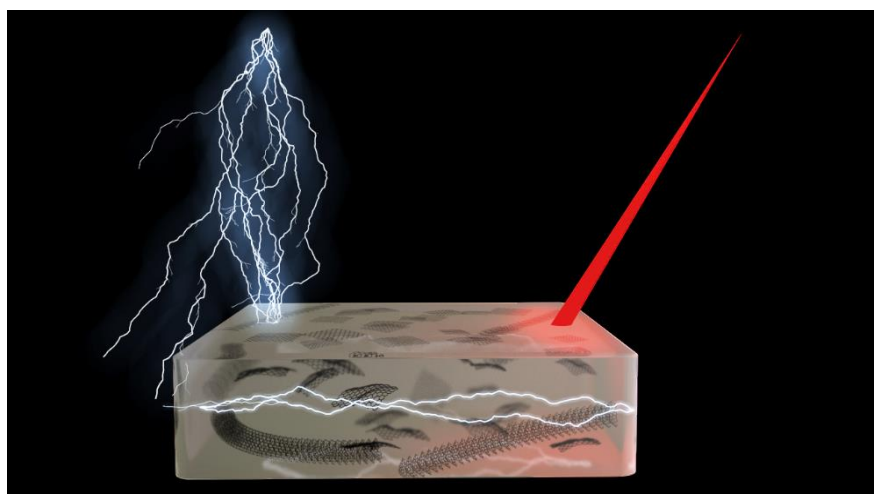

Supporting Information

# Multifunctional Biocomposites based on Polyhydroxyalkanoate and Graphene/Carbon-Nanofiber Hybrids for Electrical and Thermal Applications

*Pietro Cataldi\*, Pietro Steiner, Thomas Raine, Kailing Lin, Coskun Kocabas, Robert J. Young, Mark Bissett, Ian A. Kinloch\*, Dimitrios G. Papageorgiou\*†*

Department of Materials and National Graphene Institute, University of Manchester, Oxford Road, Manchester, M13 9PL UK
† School of Engineering and Materials Science, Queen Mary University of London, Mile End Road, London E1 4NS, UK

\* Corresponding authors e-mails: pietro.cataldi@manchester.ac.uk, d.papageorgiou@qmul.ac.uk, ian.kinloch@manchester.ac.uk

**Thermogravimetric Analysis**

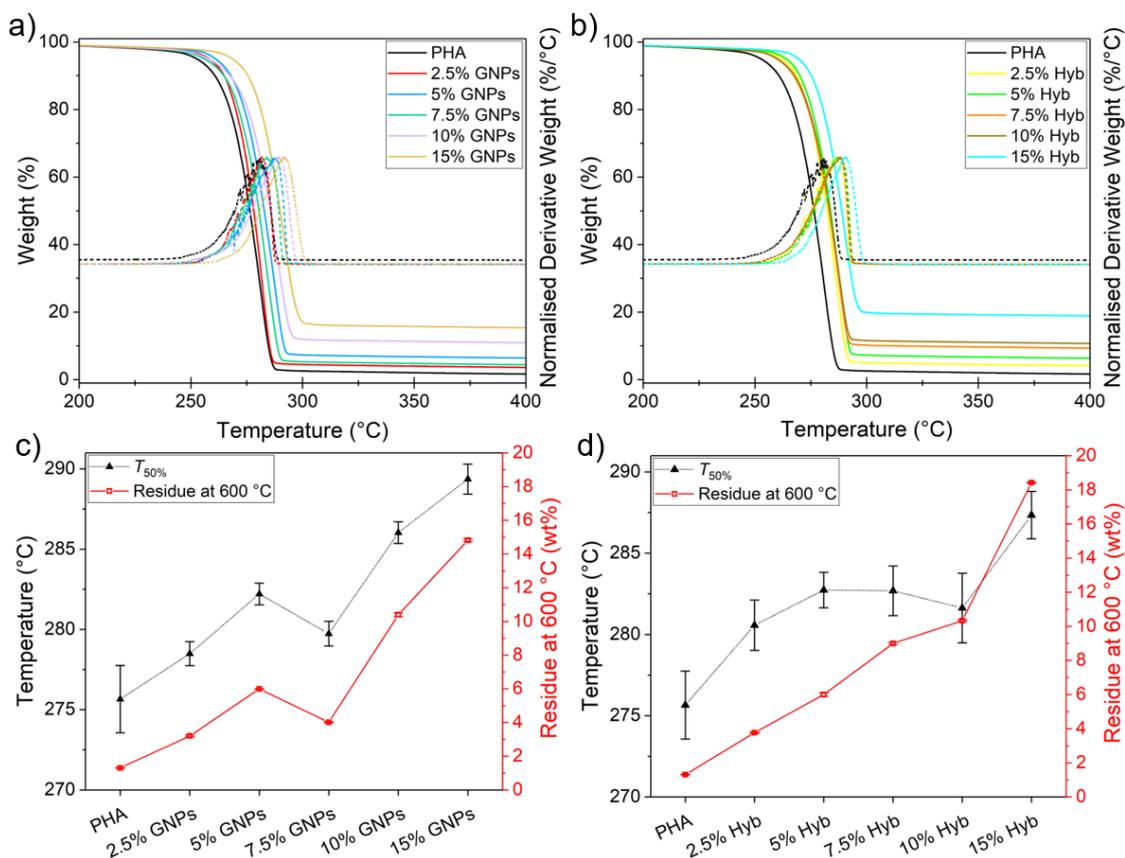

Figure S1: Thermal gravimetric analysis of the PHA nanocomposites displaying weight loss and differential thermogravimetry profiles for a) GNP-based nanocomposites and b) hybrid

nanocomposites, along with the variation of $T_{50\%}$ and residue at 600 °C for c) GNP-based nanocomposites and d) hybrid nanocomposites.

**Morphology of the Biocomposites**

The morphologies of the pure PHA sample (Figure S2a) of the 15GNP specimen (Figure S2b) and of the 15CNF nanocomposite (Figure S2c) is shown. Due to the compression moulding, the topographies of all the samples were comparable.

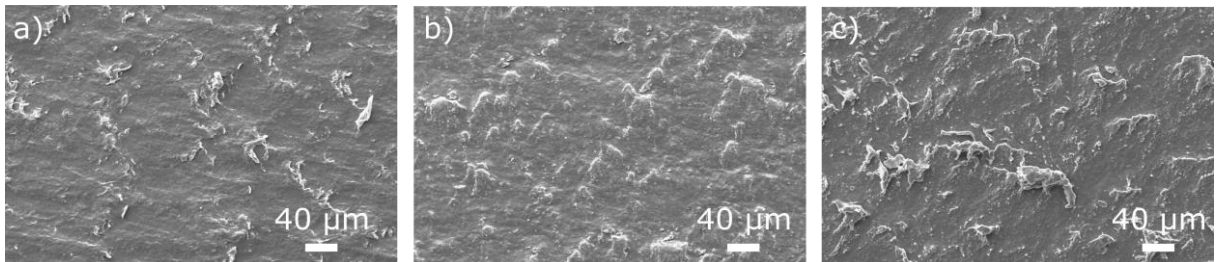

Figure S 2: SEM images of the morphology of the samples. a) pure PHA sample, b) 15GNP sample and c) 15 CNF sample.

**Cross-section of the Biocomposites**

Low magnification cross section images of the pure PHA sample (Figure S3a) of the 15GNP specimen (Figure S3b) and of the 15CNF nanocomposite (Figure S3c) are shown below. The thickness of all the samples was around 1mm as a result of the compression moulding process. Figure S3d-S3f shows high magnification SEM of the pure PHA, and of the samples containing 15 wt % of GNP and CNF, respectively. Micron-sized GNP nanoflakes are clearly observable in Figures S3e and S3f. A preferred orientation of the GNPs and of the CNFs in the direction perpendicular to the plate of the press is clearly visible in both the samples.

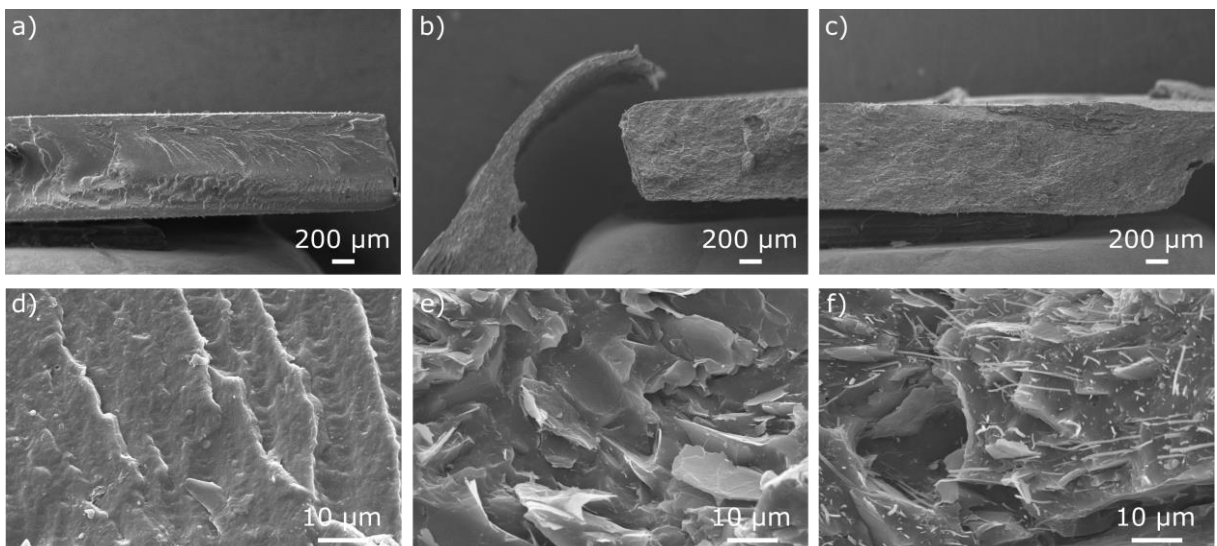

Figure S 3: SEM images of the cross section of the samples: a), d) pure PHA sample; b), e) 15GNP sample; c), f) 15 CNF sample.

## EMI Shielding Analysis

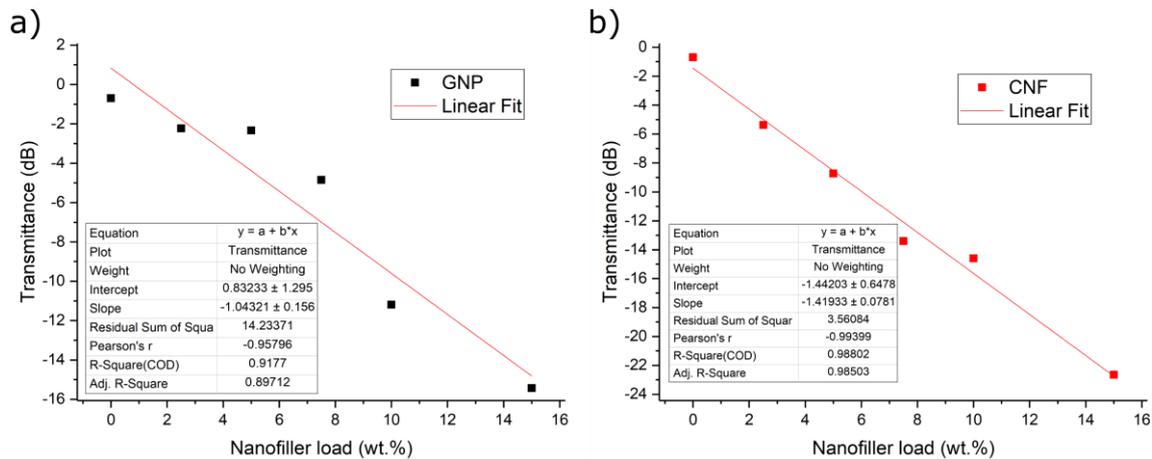

Figure S 4: a) linear fit of the transmittance as a function of GNP concentration. b) linear fit of the transmittance as a function of CNF concentration.

## Thermal Diffusivity Setup

The thermal diffusivity measurements were performed using the angstrom method. This technique use a laser source to generate periodic heat waves into the materials and a high resolution infrared camera as detector.

Plotting the temperature profile of the pixel of interest (Figure S5a) allows to measure the phase shift of the periodic wave as a function of the distance (Figure S5b). The slope of the phase over the distance is the wavenumber $k$. The thermal diffusivity is calculated using the following equation:

$$\alpha = \frac{\omega}{2k^2}$$

$\alpha = $ Thermal diffusivity

$\omega = $ Modulation frequency

$k = $ Wavenumber

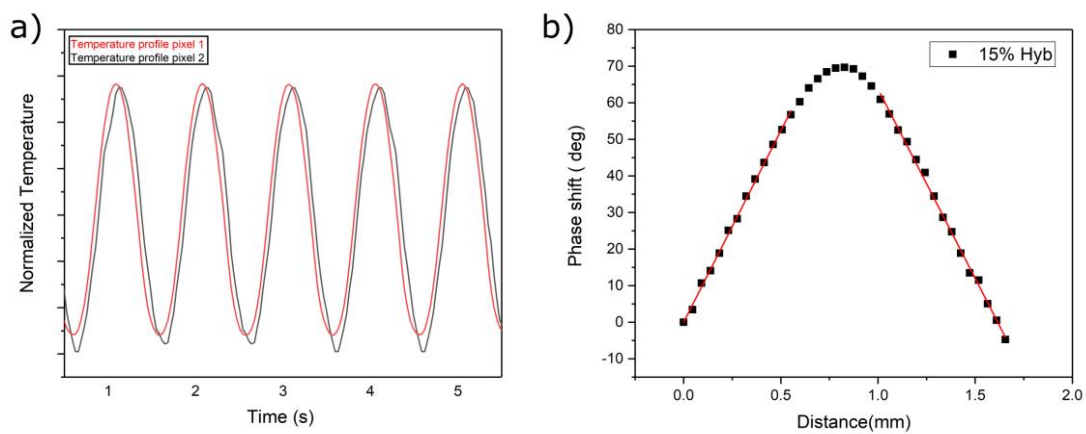

Figure S 5: a) Temperature profile for two individuals pixels, highlighting the phase shift. b) Phase shift as a function of the distance, the slope is the wavenumber.